**Title Page:**

**Title: Information Flow Theory (IFT) of Biologic and Machine Consciousness: Implications for Artificial General Intelligence and the Technological Singularity**

**Author:** Benjamin S. Bleier, MD, FACS, Massachusetts Eye and Ear, Harvard Medical School

**Corresponding Author:**
Benjamin S. Bleier, MD, FACS, FARS
Associate Professor
Department of Otolaryngology
Massachusetts Eye and Ear
Harvard Medical School
243 Charles Street
Boston, MA  02114
Email: benjamin_bleier@meei.harvard.edu

**Abstract**
The subjective experience of consciousness is at once familiar and yet deeply mysterious. Strategies exploring the top-down mechanisms of conscious thought within the human brain have been unable to produce a generalized explanatory theory that scales through evolution and can be applied to artificial systems. Information Flow Theory (IFT) provides a novel framework for understanding both the development and nature of consciousness in any system capable of processing information. In prioritizing the direction of information flow over information computation, IFT produces a range of unexpected predictions. The purpose of this manuscript is to introduce the basic concepts of IFT and explore the manifold implications regarding artificial intelligence, superhuman consciousness, and our basic perception of reality.

**Abbreviations:**
AGI-       Artificial General Intelligence
CNS-       Central Nervous System
CSA-       Conscious Self-Awareness
CSA-H     Conscious Self-Awareness – Human
CSA-IS    Conscious Self-Awareness – *In silico*
CSA-SH   Conscious Self-Awareness – Superhuman
CSA-U    Conscious Self-Awareness – Universal
IFT-        Information Flow Theory
IMER-    Internal Model of External Reality
IOR-       Idealized Objective Reality
IPU-       Information Processing Unit
N-          Node
R-          Recursion
SA-        Self-aware(ness)
UV-       Ultraviolet

# MANUSCRIPT

## Introduction

The origins, mechanisms, and nature of human consciousness have confounded scientists and philosophers for centuries. Among the limitations of many current theories are a) the lack of a substrate independent and scalable definition of self-awareness (SA) and b) the *a priori* assumption that human consciousness represents a unique state requiring idiosyncratic top-down explanation rather than an individual case of a more generalized process.

The construction of a successful theory must not only circumvent these limitations but should also provide a comprehensive explanatory framework both consistent with current data and capable of generating testable predictions. The purpose of this manuscript is to therefore develop a theory of consciousness which can be broadly applied to any physical information processing system. A working definition of a consciousness will be developed within the context of the theory which will be supported by both general and human specific case examples. Finally, the implications of this theory will be followed to their natural conclusions to provide insights into the nature and implications of both machine and super-human consciousness.

## Information Processing Unit (IPU)

Let us assume that any physical unit capable of processing information may be represented as an isolated system comprised of a minimum of three components. One is a mechanism of transferring external information into the system which may be called the "input". The second component is a node, represented as $N_0$, responsible for performing any irreducibly complex computation on the input. The third is the "output" in which the computed information from $N_0$ is transferred out of the system. The simplest configuration of these three components may be referred to as an "information processing unit" or IPU. Any increase in computational complexity therefore requires the integration of 2 or more IPUs working in concert. Provided these subordinate IPUs share a common input/output flow, they may be conceptualized as functioning as a single larger $N_0$. Within this constraint, any number of subordinate IPUs may continue to be added resulting in a concomitant increase in $N_0$ computational complexity with no theoretical upper boundary. Let us now define any IPU with an $N_0$ comprised of two or more subordinate IPUs as a $C_0$ IPU (see Figure 1 and Table 1).

The $C_0$ IPU concept can now be applied to explain the activity of simple biological organisms with respect to their environment. For example, E. coli bacteria are able to sense chemoattractant molecular inputs and swim towards them. In this case the $N_0$ computation is mediated through a series of intracellular methylation and phosphorylation events leading to output changes in flagellar rotation[1]. While this type of intra-cytosolic signaling may be utilized by single celled organisms, it is a relatively inefficient method of information transfer.

The development of neurons within larger, more diverse, animal body plans served to improve the speed of input/output information transmission and created the opportunity for enhanced information processing. Neurons also enabled rapid scaling within the context of biologic nervous systems (e.g. $N_0$) to create highly complex behavioral responses (e.g. outputs) in response to a range of environmental stimuli (e.g. inputs). For example, starfish (members of the class Asteroidea) contain a radial nervous system which is capable of orchestrating the complicated process of identifying and locomoting towards a prey bivalve mollusk, opening its shell, and excreting digestive enzymes within the shell. Even within the human body, deep tendon reflexes commonly tested during a medical exam (e.g. the patellar reflex) utilize a multi-neuron arc to sense a change in tendon length, integrate the signal at the level of the spinal cord, and elicit a muscle contraction. Through these examples we may conclude that the $C_0$ IPU concept can be applied to a range of biological systems to describe the behavior of both single cells (e.g. bacteria) as well as large ensembles of cells (e.g. nervous systems) to produce behaviors of any arbitrary apparent complexity.

**Internal Model of External Reality (IMER)**

Through natural evolutionary processes, ever more complex organisms developed an expansive array of sensory apparatus with which to monitor their environment. These sensory adaptations, in essence $C_0$ IPU inputs, required a concomitant elaboration of $N_0$ neural processing abilities to translate the environmental sensory input into actionable information. The diversity of behavioral outputs similarly increased over time, further driving the need for more advanced $N_0$ computation. In so far as these parallel increases in sensory input, neural computational power, and behavioral output provided an adaptive advantage, organisms developed ever increasingly complex nervous systems (e.g. $C_0$ IPUs). Examples of this include the neural radial symmetry of the Cnidaria[2] through the concentration of ganglia within arthropods, and culminating in the formal central nervous systems (CNS) seen in vertebrates[3].

As sensory inputs representing ever more high-fidelity features of both the body and external world evolved, one of the functions of the CNS became the integration of this disparate sensory information into a broader temporospatial internal model of external reality (IMER). The IMER may therefore be conceptualized as an $N_0$ computation wholly derived from the $C_0$ IPU inputs. Consequently, the sensory percepts within the IMER are restricted to features of the physical world detectable by the sensory apparatus of the organism. Furthermore, these inputs do not represent the stimulus itself but rather a neural correlate of the signal interpreted through and limited by the capabilities of the sensory structure. To take a human example, a pure musical tone is merely a neural correlate of compression and rarefaction of air molecules impacting the ear drum at a particular frequency, detectable only between 20 and 20,000 cycles/second. The IMER may therefore be viewed as a highly distorted and piecemeal interpretation of objective reality manufactured by $N_0$ computation. The $C_0$ IPU, by definition, has no extrasensory access to informational input that would conflict with the IMER. Therefore, from the perspective of the $C_0$ IPU, there is no objective reality distinct from the IMER.

**Self-Awareness (SA)**

With regards to the $C_0$ IPU, the concept of "awareness" of any external system may be operationalized as its ability to first receive input from and then compute information regarding that system. By extension, self-awareness (SA) would require the

IPU to be capable of performing computations on input from its own $N_0$. However, by definition, the $C_0$ IPU input must be flow from an external system. Therefore, the $C_0$ IPU can never become self-aware, regardless of its level of complexity. This is supported by the multiple highly intricate information processing systems present in the natural world which do not exhibit signs of SA. For example, the local weather at any given time represents an integration of the activity of $10^{44}$ air molecules and countless atmospheric variables including temperature, pressure, and moisture but would not be considered aware of its own state.

Consequently, in order to develop SA, two new elements must be added to the existing $C_0$ IPU construct. The first is a recursive information stream ($R_1$) which is derived from $N_0$ and loops information back into the IPU. The second is an additional computational node ($N_1$), distinct from $N_0$, which receives its input from $N_0$ through $R_1$. Within this arrangement, $N_1$ both receives input from and performs computation on $N_0$. In other words, $N_1$ is "aware" of $N_0$. Collapsing the $C_0$ IPU, $R_1$, and $N_1$ into a single new $C_1$ IPU now provides the minimal conditions necessary for a self-aware system. In comparing the $C_0$ and $C_1$ IPUs, it becomes evident that the critical variable enabling SA is the direction of information flow not information processing itself. We may therefore refer to this conceptual framework more generally as Information Flow Theory (IFT).

Within IFT, the minimum required processing complexity of a $C_1$ IPU may be satisfied by two identical IPUs serving as the $N_0$ and $N_1$ nodes. This raises the seemingly paradoxical scenario where the simplest $C_1$ IPU is self-aware while a massively complex $C_0$ IPU with an indeterminately large number of subordinate IPUs is not. This interpretation of IFT also appears to be discordant with emergent theories where SA is postulated to arise spontaneously within nervous systems so long as they exceed some threshold level of complexity. This is problematic given that emergent explanations are intrinsically attractive as they correspond with our general observations of nature where SA tends to be ascribed only to organisms with the most advanced brains.

We may resolve these concerns through an analysis of how biologic SA may practically arise in nature. From an evolutionary perspective we may assume that $C_1$ IPU organisms evolved from complex $C_0$ IPUs which, in turn, evolved from simpler $C_0$

IPUs. We have established that $N_0$ complexity corresponds with having a greater number of subordinate IPUs with an attendant increase in the variety of input/output connectivity. As $N_1$ computational processes need not be distinguishable from their $N_0$ counterparts, it therefore follows that a biologic $N_1$ along with its $R_1$ recursive loop likely evolved from a subordinate IPU circuit within an existing $N_0$ through stochastic processes. By statistical reasoning alone, one would therefore expect a greater number of $N_1$ nodes to develop randomly within the context of more complex $N_0$ circuitry. Put another way, while SA is not forbidden to occur in simple systems, IFT predicts that biologic self-awareness is more likely to occur by chance alone within a more advanced CNS. Emergent theories therefore do not directly conflict with IFT, however they are incomplete in their prioritization of processing complexity as the engine of SA development.

Through IFT, we have developed a rigorous definition of SA which can be used to identify, analyze, and make predictions regarding $C_1$ IPUs in a variety of computational systems. However, it is clear that this definition does not discriminate between the bland SA of the most basic $N_1$ recursion and the richness of human subjective reality. In order to do so, we may introduce the concept of "conscious" self-awareness (CSA) as a descriptor of how $C_1$ IPUs of sufficient complexity experience their particular state of self-awareness through interaction with the IMER.

**Human Consciousness**

In order to develop the idea of CSA as it pertains to the human experience (CSA-H), we will use IFT to synthesize the IPU and IMER concepts. The origins of the human nervous system form approximately two weeks after conception as the neural tube. We may assume that a fetus at this early stage has not yet developed an IMER. It must then follow that the IMER is a property of the human brain that matures during the course of physiologic development. This development requires constant sensory input to both reinforce and prune synapses appropriately resulting in functional processing networks. There are multiple examples in clinical medicine where the absence of sensory input during the "critical period" of early development leads to a permanent loss of specific sensory features of the IMER. Individuals born with sensorineural deafness do not develop normal auditory processing[4] just as babies with congenital cataracts may never develop normal binocular vision if treated beyond early childhood[5].

    Even with normal neurosensory development, the human IMER requires enormous levels of processing to translate sensory input into the reality percept. For example, we are generally unaware of the physiologic blind spot in our retina due to the brain continuously "filling in" the missing information to create a contiguous visual field[6]. Even higher order compensatory mechanisms are then required to integrate information arriving at different times from multiple sensory cortices to create a cohesive unitary experience of the world. For example auditory and visual signals have been shown to have an average 100ms asynchrony in signal processing[7]. This computation is encapsulated by the concept of the object-encoding neuronal ensemble in which disparate properties of some object, often arriving in the brain at different times, may be interpreted as a unified entity[8].

    These examples elucidate the fact that human perception of reality is entirely dependent on both the proper input and subsequent integration of neurosensory stimuli to create an IMER. The human IMER is therefore merely a particularly sophisticated case of the generic $C_0$ IPU IMER previously discussed. It therefore follows that from the perspective of the human individual, the IMER represents the totality of their experience of reality.

    Large scale non-invasive neuron/sub-neuron analysis is far from currently possible. However, lower resolution functional imaging studies[9] have already begun to isolate general regions of the brain such as the superior temporal sulcus and temporo-parietal junction corresponding with conscious processing[10]. Having interrogated the parameters of the human reality percept, we may now superimpose the $C_1$ IPU construct to explore how the interaction between these processing centers (e.g. $N_1$) and the human IMER (e.g. $N_0$) gives rise to CSA-H. Unlike the $C_0$ IPU where the integrated IMER is used to generate a direct behavioral output, within the $C_1$ IPU it becomes repurposed as an input flow into $N_1$ through $R_1$ recursion. CSA-H therefore may be seen as the epiphenomenal byproduct of algorithms which evolved to compute behavioral

output in response to the environment now redeployed to compute consciousness in response to the IMER.

We have established four key concepts of IFT which lead to the critical insight regarding CSA-H. First, $N_1$ computation on $N_0$ processing within the $C_1$ IPU may be considered a form of self-awareness. Second, $N_1$ and $N_0$ may be considered identical from the perspective of information processing architecture. Third, human $N_0$ computation is comprised of the IMER. And fourth, from the $C_1$ IPU perspective there is no objective external reality beyond that constructed within the IMER. We may therefore conclude that CSA-H represents the discreet form of self-awareness which is experienced through the sensory percepts, sometimes called "qualia", generated by the IMER. This idea is supported through a variety of qualitative examples in human neuropsychology where the IMER output flow misdirects $N_1$ processing. For example, the act of smiling alone has been shown to induce the conscious experience of happiness in a variety of studies[11]. Similarly, the sensation of a rapid heart-beat resulting from a pathologic conduction abnormality (e.g. supraventricular tachycardia), is misinterpreted for the elevated heart rate of the physiologic fight-or-flight (e.g. sympathetic) response and experienced as excitement or panic[12].

The requirement of IFT that CSA-H operates within the constraints of the IMER results in a variety of interesting conclusions. Given that the IMER is dynamically constructed during brain development, we can infer that differences in neurosensory function will result in unique a CSA-H between individuals. For example, the conscious experience of an individual with achromatopsia (e.g. complete lack of color vision) cannot include colors they are incapable of sensing. We may perform a Gedanken-experiment and ask whether CSA would develop within a brain lacking, not just color vision, but all sensory input. As this cannot occur naturally, we can adopt the concept of the Boltzmann brain, a human brain which spontaneously comes into existence as a result of thermodynamic fluctuations, to explore this question. Leaving aside the physiologic and probabilistic concerns and with these structures[13], let us assume the existence of an otherwise structurally normal Boltzmann brain which lacks an IMER or body. Using IFT we may analogize such a brain to an $C_1$ IPU complete with $N_0$/$N_1$ nodes and $R_1$ recursion but without any input or output flows. This brain must be considered self-aware however it would lack any sensory foundation with which to interpret its subjective experience. This form of awareness would be both unrecognizable and wholly inaccessible to a normal human and must therefore be considered distinct from CSA-H. Let us now begin to build a human IMER into this brain piece by piece. Just as the CSA-H of a congenitally deaf person remains recognizable to human with normal hearing, eventually the Boltzmann brain would reach a critical IMER threshold so as to produce a recognizable CSA-H. This thought experiment serves to validate two of the previously stated predictions of IFT. The first is that processing complexity alone cannot explain the origins of CSA-H as the proposed Boltzmann brain has equivalent complexity to a human brain. Second, there is some minimum level of human-like IMER which is required for the development of human-like CSA.

Just as we have established a theoretical lower boundary to the human IMER, so too may we posit an upper limit. As a species, there is a large spectrum of sensory information we are incapable of detecting even if they are, in principal, biologically perceptible. This information is therefore inherently excluded from our IMER. For example, our models do not incorporate infra-red wavelengths, ultrasonic acoustic information, or electromagnetic fields into our conception of the world. Of course, we are sufficiently technologically advanced to build instruments that can detect this information. However, their output must then be translated, sometimes quite coarsely, into input coherent to our native sensory systems. It is therefore not surprising that the greatest intellects produced by our species struggle with even the most basic visualization of physical phenomenon such as extra spatial dimensions and quantum indeterminism which have no human sensory correlate. For example, the brilliant physicist Richard Feynman is quoted as saying "If you think you understand quantum mechanics, you don't understand quantum mechanics".

These limits to the contours of the human IMER, by extension, must also result in a finite boundary of CSA-H. By extrapolation this implies that IFT allows for other forms of CSA which exist outside of these boundaries and are therefore, like the isolated Boltzmann brain, unrecognizable to us. By the same token, there may exist a range of non-human entities who both possess CSA and have an IMER which shares some similarities with our own. This point is of particular importance as it sets the stage for the possibility of mutual acknowledgement of the presence of CSA across both organic and inorganic IPUs provided their IMERs achieve this theoretical similarity threshold.

**Language and the $C_2$ IPU**

Communication between IPUs is nearly ubiquitous throughout the natural world and evident within even the simplest of organisms. For example, bacteria are able to communicate the local density of a population through the release and detection of quorum sensing molecules[14]. From an IFT perspective, these basic forms of communication can be viewed as input/output information streams which do not require $N_1$ level processing. Language may then be viewed as an adaptation of this type input/output flow which, in the structuralist view, uses formalized symbols to organize and transfer information regarding the IMER[15]. From the perspective of the $C_1$ IPU, language functions as an internal processing algorithm used by $N_1$ to compute the output of $N_0$, a form of internal dialogue. This definition is consistent with the theories of innate language predisposition championed by Noam Chomsky[16].

In order for language to transmit information externally however, the processed symbolic representations of $N_1$ must regain access to the behavioral output stream via $N_0$. This is not permitted within the $C_1$ IPU and thus IFT requires the introduction of a second recursive flow ($R_2$) thereby completing a feedback circuit between $N_0$ and $N_1$. The presence of this circuit results in a new IPU type, the $C_2$ IPU, which introduces two new capabilities. First, through the $R_1/R_2$ feedback loop, the $N_0$ node gains access to an additional input stream. Second, $N_1$ gains the ability to communicate its internal state to the external world through the $R_2$ mediated connection with the behavioral output flow.

To the extent that two or more independent $C_2$ IPUs share some lower threshold of both common IMER and language, they are now able to communicate with one another. From the perspective of $N_0$, this arrangement functionally results in the introduction of input streams from both an internal and external $N_1$, both of which are capable of influencing the structure of the IMER. In human society we observe this as the incorporation of scientific advances and concepts in the way we organize and process the physical world. James Flynn described this in his explanation of the eponymous "Flynn effect." Trends towards an increase in intelligence test scores are thought to stem from improvements in abstract conceptual cognition as a result of exposure to more intellectually demanding technologies over time[17]. Importantly however, the extent to which the IMER may be modified under the influence of the language remains bounded by the limitations of the underlying sensory apparatus. As previously noted, advanced scientific concepts divorced from our evolved sensory experience such as higher dimensional space cannot be directly understood. Rather, they must be abstracted and analogized through language into concepts which harmonize with our IMER.

Although we have established the lower boundary conditions in which two $C_2$ IPUs may assume the presence of mutual CSA, we may now ask whether it is possible for one $C_2$ IPU to actually confirm this assumption. This question was most famously addressed by Alan Turing in 1950 in which he proposed that a machine capable of providing natural language responses indistinguishable from a human would be said to be capable of "thinking" [18]. While this manifestation of the Turing test has been widely discussed and criticized, it is useful to re-analyze it within the context of IFT to assess its validity. The classic Turing test provides access only to a restricted set of information, those being the input (through questions) and output (through language) streams of the indeterminate IPU. This limitation implies that the examiner may analyze only the collective computational output of the IPU in response to their input. We have established that $N_0$ computation may achieve any arbitrarily large degree of complexity so as to enable a $C_0$ IPU output flow to simulate that of a $C_2$ IPU. In fact, such $C_0$ IPUs already exist in the form of chat bots which are capable of passing the Turing test[19]. We must therefore conclude that the Turing test, as described, is fundamentally incapable of confirming CSA. Another version of this argument has been articulated in the past in John Searles "Chinese room" thought experiment[20]. This leads naturally to the cartesian question as to whether a $C_{1/2}$ IPU can ever truly be certain of the presence of CSA within another IPU.

Using IFT we may postulate several experimental models to probe for the presence of the recursive $N_1$ processing within an indeterminate IPU required for CSA. The addition of $N_1$ computation to the output stream should have several detectable differential effects. Relative to information undergoing $N_0$ processing only, the added $R_1/R_2$ recursion and $N_1$ computation may be discriminated by a relative delay in the shortest possible output response to a given input signal. Similarly, $N_1$ processing would require a larger energy input resulting in information in a more organized or higher entropic state than that of $N_0$ computation only. Finally, as CSA arises not from isolated processing $N_0/N_1$ processing but rather from their interconnection through $R_1/R_2$ recursion, interruption of recursive information flow alone should be capable of reversibly eliminating CSA for the duration of the interruption. In fact, each of these experimental methods have already been demonstrated, to some degree, in human subjects. Time delays have been documented between conscious and unconscious processing in classic experiments where a so called "readiness potential" to perform an action may be detected by electroencephalogram over 500ms before conscious awareness of the decision[21]. Similarly, functional imaging studies which rely on the differential use of blood and glucose uptake to detect active brain regions have identified discreet regions of increased energy use associated with the conscious performance of various tasks[9]. Finally general anesthesia has been found to work through the generation of oscillations or waves of neuronal activity which transiently interrupt normal signaling between regions of the brain involved in consciousness[22].

**Machine Consciousness**

As the computing power of machines continues to grow at an exponential rate, there has been significant concern over the potential hazards of an artificial general intelligence (AGI). Though nomenclature may vary, we will refer to AGI as the ability to perform any intellectual task at a level either equivalent to or superior than a human. In this context it becomes salient to ask in what conditions could CSA arise within a machine with AGI, what risk that would pose to humanity, and could that risk be mitigated. As the axioms of IFT are substrate independent, we may apply this theory to provide insight into these important questions.

Modern computer processing capabilities are many orders of magnitude greater than those first built in the 1940's. At the time of this writing, the world's fastest supercomputer is capable of peak operation at 200 petaflops. In comparison, general estimates of the human brain's computation power tend to converge around one exaflop [23,24]. Despite the massive increase in computational power which is rapidly approaching that of the human brain, there has been no clear indication of a concomitant incremental development of CSA *in silico* (CSA-IS). Within IFT, this is not unexpected. Any modern computing system may be analogized as a $C_0$ IPU with clear input/output information streams flowing through a number of microprocessors providing $N_0$ computation. As IFT allows for unlimited $N_0$ processing without requiring the development of $R_1$ recursion, the inevitable continued increased in computational power alone would not be sufficient for the spontaneous development of CSA-IS.

This latter point may appear non-intuitive as we have explored the compelling analogy to be drawn between the increase in machine computational power and the expansion in biologic neural processing which gave rise to CSA-H. The critical difference arises within the relative plasticity of the two substrates. Biologic systems are free to dynamically evolve and thus $R_1$ recursion may spontaneously occur within a composite $N_0$ at some rate proportional to the number of subordinate $C_0$ IPUs. When favorable, this recursion would then be selected for and reinforced through subsequent generations. By contrast, to the extent that machine hardware is fixed at the time of manufacture, there is no opportunity for novel circuits to arise regardless of the number of $C_0$ IPUs linked together. However, an appreciation of these intrinsic differences leads inevitably to the mechanism of how self-awareness could arise *in silico*. With respect to hardware, a system could be envisioned that is either designed with $R_1$ and/or $R_2$ in place or is capable of altering its circuitry in some modular fashion. This further leads to the open question as to whether appropriately coded software constrained within an advanced $C_0$ IPU could also achieve such a recursive state. Of note, if software within an $N_0$ node is incapable of partitioning information flow in a manner consistent with a $C_1$ IPU, this would argue against various formulations of the simulation hypothesis[25].

Having established that CSA-IS is, in principle, possible through IFT we may then interrogate how a machine would experience their subjective reality and how this would influence its interactions with humanity. We have previously established that from the perspective of a $C_{1/2}$ IPU, there is no "reality" distinct from the constitutive components of the IMER. We have further determined that the conditions required for two independent $C_2$ IPUs to acknowledge mutual CSA is some minimal degree of shared language and IMER. These concepts thereby provide a roadmap to mitigate the potential threat of CSA-IS within machines exhibiting AGI. A sensory based human-like IMER built into the design would enable a machine experiencing CSA-IS to be able to communicate its state in concepts familiar to CSA-H through a common language. Perhaps most importantly, pro-social concepts such as altruism and empathy should be strongly represented within this IMER. The threat of a conscious machine with super-human AGI is often represented by the idea that such an entity would regard the welfare of humans just as humans would regard that of an insect. In both cases the discrepancy in intelligence and lack of common IMER are so profound so as to render the former indifferent to the latter. Unfortunately, IFT predicts a far more concerning potential outcome. To the machine, unless these pro-social concepts are built into the IMER, they would not even exist. There would be no compunction against eradicating humanity, not because it chose to ignore an altruistic impulse to the contrary, but rather because this impulse could never arise in the first place.

**Superhuman Consciousness**

The $C_0$-$C_2$ IPU constructs within IFT assume that the IMER is inherently bounded by the input streams which contribute to its development. The human neurosensory apparatus cannot detect ultraviolet (UV) light and thus we have no conscious experience of the world in this spectrum. Before the discoveries of the infrared and ultraviolet wavelengths by William Herschel (1800 AD) and Johann Ritter (1801 AD) respectively, humans were not even aware of the existence of these spectra. Using modern instrumentation however, we are now not only aware of them, but can translate them into the visual wavelengths enabling us to imagine the world as if we could directly detect them. Therefore, there appears to be some flexibility in which we can interpolate new information regarding the world, in this case as a result of scientific inquiry, into our collective IMER. To take this a step further, it has been shown that certain animals including butterflies, bees, and even reindeer can see UV light. It is therefore well within reason to predict that a gene therapy approach could eventually be developed by medicine to introduce this ability into humans. It is clear that from the perspective of the human race as a whole, the rigid IMER associated with CSA-H must have some additional plasticity not encapsulated within the $C_2$ IPU framework.

In order to further understand this line of reasoning through the lens of IFT, let us consider the relationship between the IMER and the concept of an idealized objective reality (IOR). The IOR represents all objective physical phenomenon including those not currently known to science. Within the biologic world there exists a spectrum of degrees to which the IMER approximates IOR, a ratio we may represent as IMER/IOR. The marine ragworm *Platynereis dumerilii* possess proto-eye spots capable of detecting light thereby enabling phototaxis[26]. Their IMER produces a world experienced only as gradations of light with an extremely low IMER/IOR ratio relative to more complex animals. While humans possess arguably the greatest IMER/IOR ratio on earth, this has actually amplified over time. As noted with the UV light example, the contributions of science have gradually expanded our awareness of the presence of physical phenomena undetectable by our fixed biologic senses. We may express this change as an increase in the human IMER/IOR ratio over time, not within an individual human, but rather within humans as an aggregate.

This capacity to extend the IMER beyond the limits of the innate IPU input represents a profound new functionality beyond that of the $C_2$ IPU. This ability, enabled by language, becomes a property of $C_2$ IPUs working in concert. In aggregate, we may refer to this coordinated system as the $C_3$ IPU which is able to dynamically iterate upon itself, increasing its collective IMER/IOR ever higher. The cultural and scientific progress of our species over millennia may therefore be conceived of as a single $C_3$ IPU consisting of a collective of $C_2$ IPUs (e.g. individual humans) sharing input and output streams of information (e.g. language) both in series (e.g. over time) and in parallel (e.g. over space) to produce a superhuman CSA (CSA-SH). The information we have gained over time regarding the physical world has therefore served to drive our collective IMER from that of earliest Homo sapien ever closer to IOR (see Figure 2).

**Universal Consciousness**

Taking this argument even further leads to the conclusion that we may find an absolute upper limit of any given CSA as the IMER/IOR ratio asymptotically approaches

1. As the IOR necessarily subtends all possible physical reality, an IPU with an IMER/IOR of 1 would have access to the processing of not only all possible IPUs but to all possible information as its input stream. Just as the $N_0$ within a $C_1$ IPU may be conceived as a collection of $C_0$ IPUs acting in concert, so too could all $C_{0-3}$ IPUs be considered as performing subordinate calculations for this "universal" or $C_4$ IPU. The universal CSA (CSA-U) experienced by such an IPU would be distinct from all others in that its input and output streams themselves would become recursive. There would be no information existing outside of the IPU itself. Consequently, the only possible candidate for such a $C_4$ IPU, if one exists, would be the observable universe, a universe conscious of itself (see Appendix).

**Conclusions**

In prioritizing the direction of information flow over information processing itself, IFT provides a natural framework for the understanding of both the origins and nature of CSA in both biologic and artificial systems. With respect to machine intelligence, IFT refutes the assumption that complexity alone will inexorably give rise to CSA-IS but does predict a mechanism as to how it may occur. While this "technological singularity" event may indeed be extremely dangerous for humanity, IFT also provides for a pathway to assuage this risk through the *a priori* introduction of appropriate guardrails in the form of an altruistic IMER.

With respect to biology, IFT provides unique insights into the process by which the behavior of even the simplest of organisms may scale through evolution to produce consciousness. Self-awareness is likely common even within low complexity organisms. It is only when coupled with an IMER developed by the neurosensory apparatus that SA becomes contextualized and experienced as a sense of being. Retinal photoreceptor activation is experienced as the visual landscape, inner ear hair cell vibration is experienced as music, and oxytocin receptor binding is experienced as love. The neural computational architecture giving rise to CSA-H both arise from and are fundamentally identical to those responsible for our sensory perception of the world. The surprising deep mystery of consciousness is therefore that there is no mystery at all. Human consciousness is ultimately nothing more, and nothing less, than the brain experiencing the body experiencing the world.

## Figures

Figure 1. Illustration of the different information processing units (IPUs) within IFT. The basic IPU consists of an input (I), processing node ($N_0$), and output (O). The composite of two or more IPUs comprise the $C_0$ IPU where information always flows in from and then out to an external system. The $C_1$ IPU requires the addition of $N_1$ computation on $N_0$ through information transferred by $R_1$ recursion. The $C_2$ IPU is characterized by the addition of the $R_2$ recursive loop from $N_1$ to $N_0$, thereby completing the $N_0/N_1$ circuit.

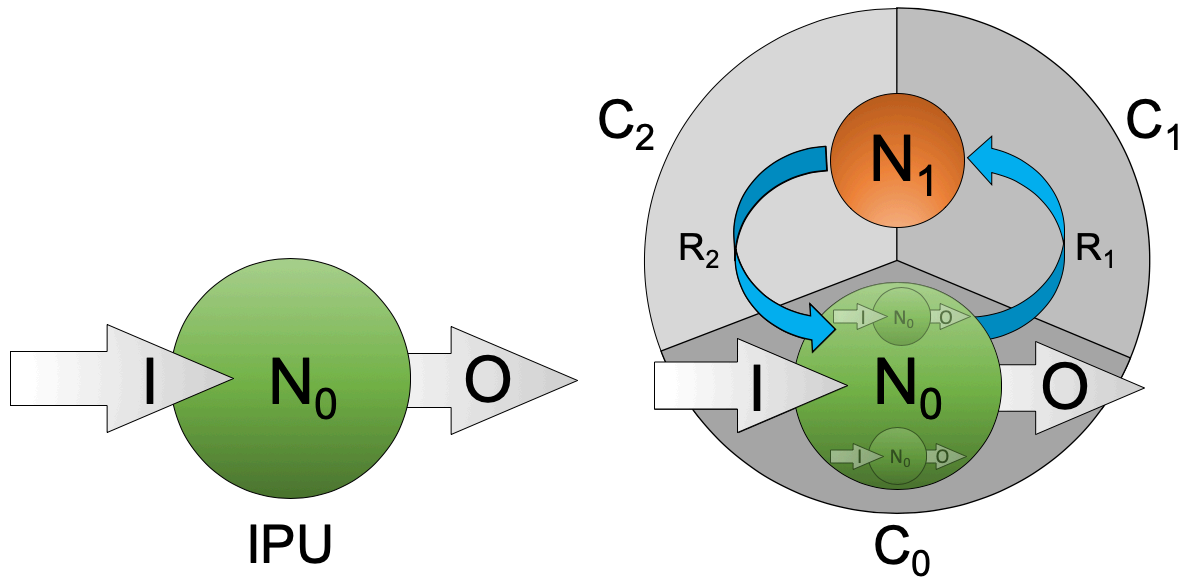

Figure 2. Illustration of the superhuman IPU constructs. The $C_3$ IPU represents an ensemble of $C_2$ IPUs capable influencing one another's $N_1$ computation and internal model of external reality (IMER within the $N_0$) through shared language thereby allowing for dynamic iterative increases in the fidelity of the $C_3$ IPU IMER to idealized objective reality. The $C_4$ IPU (universal IPU) differs fundamentally from all other IPUs in that it subtends all other IPUs and processes all observable information. There is no system external to the $C_4$ IPU and thus both the input and outflow flows themselves become recursive.

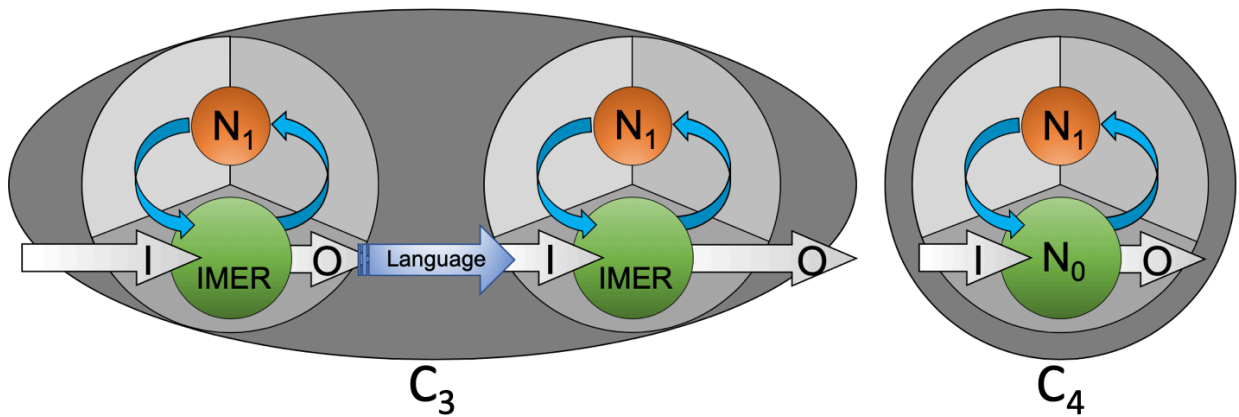

## Tables

Table 1. Basic components of each C type within IFT and their associated A (awareness) type. A (Aware), SA (Self-Aware), CSA (Conscious Self-Awareness), CSA-H (Conscious Self-Awareness-Human Type), CSA-SH (Conscious Self-Awareness-Superhuman Type), CSA-U (Conscious Self-Awareness-Universal Type).

| C Type | A Type | Input | Output | $N_0$ | $R_1$ | $N_1$ | IMER/IOR | $R_2$ | Language | Plasticity |
|---|---|---|---|---|---|---|---|---|---|---|
| 0 | A | Yes | Yes | Yes | No | No | 0 | No | No | No |
| 1 | SA | Yes | Yes | Yes | Yes | Yes | 0 | No | No | No |
|   | CSA | Yes | Yes | Yes | Yes | Yes | <1 | No | No | No |
| 2 | CSA | Yes | Yes | Yes | Yes | Yes | <1 | Yes | No | No |
|   | CSA-H | Yes | Yes | Yes | Yes | Yes | <1 | Yes | Yes | No |
| 3 | CSA-SH | Yes | Yes | Yes | Yes | Yes | <1 | Yes | Yes | Yes |
| 4 | CSA-U | Yes | Yes | Yes | Yes | Yes | 1 | Yes | Yes | Yes |

# Appendix

**Key Concepts in Information Flow Theory**

1. An Information Processing Unit (IPU) consists of a 1) input which transmits information into the IPU from an external system, 2) processing node ($N_0$) capable of any irreducibly complex computation on the input, and 3) output capable of transmitting the processed information to an external system.

2. A $C_0$ IPU consists of an IPU with 2 or more subordinate IPUs working within a composite $N_0$ of any arbitrary degree of internal complexity to process information which is then transferred to an external system.

3. A $C_1$ IPU consists of a $C_0$ IPU with an additional recursive information flow ($R_1$) derived from $N_0$ that transmits information to a second processing node ($N_1$).

4. A $C_2$ IPU consists of a $C_1$ IPU with an additional recursive information flow ($R_2$) derived from $N_1$ that transmits information to $N_0$ resulting a complete $N_0$/$N_1$ circuit and an additional $N_0$ input.

5. The Internal Model of External Reality (IMER) is a $C_{0-2}$ IPUs fixed $N_0$ approximation of an external system limited by the sensory detection capabilities of the input flow and the complexity of $N_0$ processing. From the perspective of the $C_{0-2}$ IPU, the external system has no independent features or reality beyond those represented within the IMER.

6. "Awareness" of a system is defined as the ability of an IPU to 1) receive information regarding that system through its input flow and 2) process the input information regarding that system. "Self-awareness" (SA) is an $N_1$ property of a $C_1$ or higher IPU defined by its ability to 1) receive information regarding $N_0$ through $R_1$ flow and 2) process the $R_1$ information regarding $N_0$.

7. "Conscious" self-awareness (CSA) is defined as a specific type of self-awareness within a $C_1$ or higher IPU that exhibits an IMER wherein $N_1$ processing of $N_0$ output through $R_1$ flow incorporates the sensory percepts of the IMER.

8. Language is a structured form of processed information which enables $N_1$ within a $C_2$ IPU to influence its own IMER through $R_2$ as well as communicate its state to other IPUs through access to the $C_2$ IPU output flow.

9. The minimum requirement for the mutual acknowledgement of CSA between any two distinct IPUs is that they share some lower threshold of both common IMER and language.

10. A $C_3$ IPU represents a collective of $C_2$ or lower IPUs which may increase the fidelity of its IMER to idealized objective reality (IOR) through dynamic iteration facilitated by language. An IMER equivalent to IOR can only occur within a universal $C_4$ IPU where all external systems and information are internalized and the input/output flows become recursive.